# Cavities as a source of outbursts from comets

**Sergei I. Ipatov**[1,2]

[1]Catholic University of America, Washington, DC, USA, siipatov@hotmail.com,
[2]Space Research Institute, Moscow, Russia

**ABSTRACT**

Analysis of observations of natural and triggered outbursts from different comets testifies in favor of existence of large cavities with material under gas pressure below a considerable fraction of a comet's surface. Based on analysis of images of the cloud of material ejected from Comet 9P/Tempel 1 after the collision of the Deep Impact (DI) module with the comet, we studied the time variations in the rate and velocities of ejection of observed particles (mainly icy particles with diameter $d$<3 μm). Comparison of these dependencies with the theoretical dependencies allowed us to understand the time variations in the rate of the outburst triggered by the impact. The latter variations testify in favor of that there were cavities at the place of DI ejection. The beginning of the increase of the main outburst at 8 s after the DI impact could be caused by excavation of a relatively large cavity that contained dust and gas under pressure. The upper boarder of the cavity could be located at about 5-10 meters below the surface of the comet. This cavity could be deep because the excavation from the cavity could last for at least a few tens of seconds. With the increase of the crater, more cavities could be excavated. The outburst decreased at ~60 s after the impact. Besides the 'fast' outburst caused by ejection from cavities, there was a 'slow' outburst ejection, which was similar to the ejection from a 'fresh' surface of a comet and could be noticeable during 30-60 min. The 'fast' outburst with velocities ~100 m/s probably could continue for at least several tens of seconds, and it could significantly increase the fraction of particles ejected with velocities ~100 m/s compared with theoretical models of ejection.

**INTRODUCTION**

In 2005 the impact module of the Deep Impact (DI) spacecraft collided with Comet 9P/Tempel 1 (A'Hearn et al. 2005). The outburst triggered by this impact was one of many other outbursts from comets. The DI spacecraft observed natural outbursts from Comet Tempel 1. Observed outbursts from different comets testify in favor of the existence of relatively large cavities with dust and gas under pressure.

In this paper, I discuss observed outbursts from different comets and the internal processes (e.g. crystallization of amorphous ice and sublimation of CO ice) that can cause the outbursts (Section 1), the cometary-like activity of objects moving in typical asteroid orbits (Section 2), the outburst triggered by the DI impact (Section 3), peculiarities of the DI ejection and cavities in Comet Tempel 1 (Section 4).

**1. Observed outbursts from different comets**

Astronomers observed outbursts from several comets. Cometary activity was observed even for some comets (e.g. Comet Hale-Bopp) moved outside of Jupiter's orbit and for some objects moving in typical asteroid orbits. Natural outbursts from comets could last for weeks or months.

The total mass of material ejected at the 2007 October 24 outburst of Comet 17P/Holmes (~1-4% of the nucleus mass of the comet, i.e. (1-3)×$10^{11}$ kg) was much greater than that at the DI collision (~$10^7$ kg). Schleicher (2009) concluded that production of OH decreased by a factor of 200-300 during 124 days after the outburst of Comet 17P/Holmes in 2007, but it was still greater than before the outburst. It shows that the ejection of material from a 'fresh' surface of a comet can make a noticeable contribution to the total ejection from the comet for many days. Schleicher (2009) suggested that the explosion occurred at greater depth in Holmes than in other comets. Possibly, the explosion at such depth can explain large (up to 125 m/s) observed on-sky velocities of 16 large (with effective radii between ~10 and ~100 m) fragments of Comet 17P/Holmes (Stevenson et al. 2010).

Paganini et al. (2010) observed the outburst activity of Comet 73P/Schwassmann-Wachmann 3 in 2006 May. They obtained a decrease in gaseous productivity of this comet by a factor of 2 in about a week. Prialnik et al. (2004) concluded that the outburst of Comet 1P/Halley could take place during a few months when the comet moved at a distance greater than 5 AU from the Sun. Results obtained by Ivanova et al. (2011) support the idea that the observed activity of Comet 29P/Schwassmann-Wachmann 1 at distance >5.5 AU requires a permanent demolition of the upper surface layers.

It is considered that outbursts are mainly caused by internal processes. Internal gas pressure is considered to be one of the main reasons of splitting of comets (Boehnhardt 2004, Fernandez 2009). Boehnhardt (2002) concluded that if the gas pressure cannot be released through surface activity, the tensile strength of the nucleus material can be exceeded and fragmentation of the comet occurs. Mechanism of activity of comets (including Comet Tempel 1) was studied by Belton et al. (2007) and Belton (2010). Belton et al. (2008) concluded that natural outbursts on Comet 9P/Tempel 1 were caused by that at some depth the stress of gas overwhelmed the strength and overburden pressure of cometary material. In their opinion, the events might be triggered by changing thermal stresses or other processes in surface material in response to a cooling of the surface.

Comet nuclei are assumed to be of porous structure. For example, A'Hearn et al. (2005) and Richardson et al. (2007) considered that the bulk density of Comet Temple 1 is ~0.6 and ~0.4 g/cm$^3$, respectively. Porosity also testifies in favor of existence of cavities. Sources of gas that can fill cavities and pores in comets include the crystallization of amorphous ice and the sublimation at 'internal' surfaces (Möhlmann 2002).

Prialnik and Bar-Nun (1987) and Prialnik et al. (2004) supposed that crystallization of amorphous ice in the interior of the porous nucleus, at depths of a few tens of meters, caused the release of gas. Prialnik et al. (2008) noted that since amorphous ice would only be presented below the surface, there is little direct evidence for the amorphous state of cometary ice. Davies et al. (1997) presented spectra of Comet Hale-Bopp (C/1995 O1) when the comet was 7 AU from the Sun, and they concluded that the absence of the 1.65-µm absorption feature of crystalline ice suggests that the cometary ice was probably in an amorphous state at the time of these observations. A similar conclusion was made by Kawakita et al. (2004) based on studies of spectra of Comet C/2002 T7 (LINEAR) at 3.52 AU from the Sun. The role of crystallization of amorphous ice in bursts of comet activity was discussed in several other papers. A few references and examples of such bursts are presented by Prialnik (2002) and Belton (2010).

Ishiguro et al. (2010) concluded that the 2007 outburst of Comet 17P/Holmes was caused by an endogenic energy source. Reach et al. (2010) supposed that the explosion of Comet 17P/Holmes was due to crystallization and release of volatiles from interior amorphous ice within a subsurface cavity: once the pressure in the cavity exceeded the surface strength, the material above the cavity was propelled from the comet.

Based on millimeter-wavelength continuum observations, Altenhoff et al. (2009) suggested that the recent "spectacle" of Comet 17P/Holmes can be explained by a thick, air-tight dust cover and the effects of $H_2O$ sublimation, which started when the comet arrived at the heliocentric distance ≤2.5 AU. The porous structure inside the nucleus provided enough surface for additional sublimation, which eventually led to the break up of the dust cover and to the observed outburst. Gronkowski and Sacharczuk (2010) suggested that it is questionable that the amorphous water ice survived up to now in the nucleus of 17P/Holmes because it belongs to Jupiter-family comets and such comets should after several hundred orbits convert all of the water ice in the nucleus into a crystalline form.

Huebner (2008) paid attention that amorphous water ice or clathrate hydrates have not been detected in interstellar clouds, star-forming regions, or outer solar system bodies. He supposed that it is quite possible that amorphous ice cannot survive very long on surfaces and that its existence in comet nuclei still needs to be proven. Belton (2010) noted that there is no observational confirmation of the presence of amorphous ice in cometary nuclei. He considered the search for amorphous ice in the interior of Jupiter-family comets to be the most significant objective for future space missions.

Schleicher (2009) suggested that, besides water ice sublimation or a subsequent phase change in the ice, an alternate source of a buildup in pressure could be sublimation of a more volatile ice such as CO or $CO_2$ at a lower temperature than required for water ice. Several other potential mechanisms of outbursts are also discussed: (a) the polymerization of hydrogen cyanide HCN; (b) thermal stresses; (c) and anneling of the amorphous water ice; (d) meteoritic impacts. References to the papers that considered such mechanisms can be found in (Gronkowski & Sacharczuk 2010, Ivanova et al. 2011).

Kossacki and Szutowicz (2011) made calculations for several models of the explosion of Comet 17P/Holmes. They concluded that the nonuniform crystallization of amorphous water ice itself is probably not sufficient for an explosion, which could be caused by a rapid sublimation of the CO ice leading to the rise of gas pressure above the tensile strength of the nucleus. In their models, the initial sublimation front of the CO ice was located at a depth of 4 m, 10 m, or 20 m. It was shown that the pressure of CO vapor can rise to the threshold value only when the nucleus is composed of very fine grains (of diameter of a few microns). Gortsas et al. (2011) also did not include amorphous ice in their models of the activity of Comet C/1995 O1 (Hale-Bopp). They considered sublimation of water and CO ice not only at the surface, but also from interior.

Based on analysis of pre-Deep Impact images obtained at the Indian Astronomical Observatory, Vasundhara (2009) concluded that from some active regions the grains were ejected from Comet Tempel 1 with a velocity distribution with an upper limit of 70 m/s and from a broad region they were ejected with an upper limit of 24 m/s. According to Feldman et al. (2007), at the June 14, 2005 natural outburst from this comet, velocities of ejection were 60-145 m/s. Sarugaku et al. (2010) obtained that the dust cloud caused by the outburst from Comet 217P/LINEAR expanded at a velocity of 120-140 m/s. Ejection velocities of outburst particles of Comet 29P/Schwassmann-Wachmann 1 were about 250±80 m/s (Trigo-Rodriguez et al. 2010).

## 2. Cometary-like activity of objects moving in typical asteroid orbits

A few Main Belt Comets, or "icy-asteroids", were found in the asteroid belt (Bertini 2011). In our opinion, cometary activity of asteroid 7968 Elst-Pizarro, also known as Comet 133P/Elst-Pizarro, ($a$=3.16 AU, $e$=0.16, and $i$=1.39$^o$) could be caused by the same internal processes as natural and triggered outbursts from other comets (including Comet Tempel 1), but its solid crust could be thicker than that of Comet Temple 1. In 1996, 2002, and 2007, the object Elst-Pizarro had a comet tail for several months. The asteroid orbit of this object is stable (Ipatov & Hahn 1999). Based on studies of the orbital evolution of Jupiter-crossing objects (Ipatov & Mather 2003, 2004), Ipatov and Mather (2007) supposed that the object Elst-Pizarro earlier could be a Jupiter-family comet, and it could circulate its orbit also due to non-gravitational forces.

Hsieh et al. (2010) concluded that activity of Comet 133P/Elst-Pizarro was consistent with seasonal activity modulation and took place during hemisphere's summer, when the comet received enough heating to drive sublimation. We suppose that there could be natural outbursts during the 'summer', and they could be one of the sources of observed activity of the comet. It could be possible that vaporized material formed under the crust moved outside through narrow holes for a long time. There can be a lot of ice under the crust of the object Elst-Pizarro, and this ice produced a comet tail after the crust had been damaged in some way (e.g. due to high internal pressure).

Cometary-like activity was also observed for P/2010 A2 (LINEAR), which has a typical asteroid orbit ($a$=2.29 AU, $e$=0.12, and $i$=5.26°). The total amount of the dust released from this object during eight months was estimated by Moreno et al. (2010) to represent 0.3% of the nucleus mass. They supposed that some subsurface ice layer exists in this object. Several other authors (e.g. Jewitt et al. 2010, Snodgrass et al. 2010) believe that the trail of P/2010 A2 is the result of the collision between two asteroids, not of cometary activity, because this object is close to the inner edge of the asteroid belt. In our opinion, if this object contains ice (e.g. it was captured from a comet's orbit), then the internal gas pressure could also play a role in the ejection of particles from this object, but this role should not be considerable because typical velocities of ejection from cavities under gas pressure are greater than the velocities (<1 m/s) obtained by Jewitt et al. (2010).

## 3. Triggered Deep Impact outburst

Based on analysis of images of the cloud of ejected material made by the DI cameras during the first 13 min after the collision of the DI impact module with Comet Tempel 1, Ipatov and A'Hearn (2010, 2011) and Ipatov (2011) studied the time variations in the rate and velocities of ejection of observed particles (mainly with diameter $d$<3 μm). These variations differed from those for the model based on theoretical studies of impact events and testify in favor of that there were cavities (with gas and dust under pressure) inside the comet at the place of ejection.

Analysis of maxima or minima of plots of the time variations in distances $R$ of contours of constant brightness from the place of ejection (at $R$>1 km, i.e. outside of regions of saturated pixels) allowed Ipatov and A'Hearn (2011) to estimate the characteristic velocities of particles (at a distance of a few km from the place of ejection) at several moments in time $t_e$ of ejection after impact for $t_e$≤115 s. Other approaches for estimates of the velocities were also used. For example, characteristic values of projections $v_p$ of velocities (onto the plane perpendicular to the line of sight) of observed particles were about 7 km/s at $t_e$~0.2 s, $v_p$≈250 m/s at $t_e$≈4 s, $v_p$≈100 m/s at $t_e$~10-20 s, and $v_p$≈20-25 m/s at $t_e$~70-115 s. At 1<$t_e$<100 s the time variations in $v_p$ can be approximately considered to be proportional to $t_e^{-\alpha}$ with $\alpha$~0.7-0.75.

Analysis of time variations in the size of the bright region of ejected material allowed us to estimate the time variations in the relative amount of observed ejected particles. There was a local maximum of the rate of ejection at $t_e$~10 s with $v_p$~100 m/s. At the same time, the considerable excessive ejection in a few directions (rays of ejecta) began, there was a local increase in brightness of the brightest pixel, and the direction from the place of ejection to the brightest pixel quickly changed by about 50°. In images made during the first 12 s and after the first 60 s, this direction was mainly close to the direction of the impact. Between 8 and 60 seconds after the impact, more small bright particles were ejected than expected from crater excavation alone. An outburst triggered by the impact could cause such a difference. The sharp (by a factor of 1.6) decrease in the rate of ejection at 55<$t_e$<72 s could be caused by a decrease in the outburst that began at 8-10 s.

For the model *VExp* with $v_p$ proportional to $t_e^{-\alpha}$ at any $t_e$>1 s, the fractions of observed (not all) material ejected (at $t_e$≤6 and $t_e$≤15 s) with $v_p$≥200 and $v_p$≥100 m/s were estimated to be about 0.1-0.15 and 0.2-0.25, respectively, if we consider only material observed during the first 13 minutes. The 'fast' outburst with velocities ~100 m/s probably could last for at least several tens of seconds, and it could significantly increase the fraction of particles ejected with velocities ~100 m/s, compared with the estimates for the model *VExp* and for the normal ejection. The above estimates are in accordance with the estimates (100-200 m/s) of the projection of velocity of the leading edge of the DI dust cloud made by several observers and based on various ground-based observations and observations made by space telescopes (the review of such papers is given by Ipatov and A'Hearn (2011)). Velocities of about 100-200 m/s were observed at natural outbursts from other comets (see the end of Section 1).

The excess ejection of material in a few directions (rays of ejected material) was considerable during the first 100 s, and it was still observed in images made at $t$~500-770 s. This finding shows that the 'fast' outburst could continue at $t_e$~10 min. The sharpest rays were caused by material ejected at $t_e$~20 s. In particular, there were excessive ejections, especially in images made at $t$~25-50 s after impact, in directions perpendicular to the direction of impact. Directions of excessive ejection could vary with time.

Our studies did not allow us to estimate accurately when the end of ejection occurred, but they do not contradict a continuous ejection of material during at least the first 10 minutes after the collision. The duration of the outburst could be longer than that of the normal ejection, which could last only a few minutes. Besides the 'fast' outburst caused by ejection from the cavities, there was a 'slow' outburst ejection, which was similar to the ejection from a 'fresh' surface of a comet and could be not very small at $t_e$~30-60 min. According to Cochran et al. (2007), there was no considerable fragmentation of icy grains that increased the brightness of the cloud (for the same total mass of the cloud). As the total brightness of the DI cloud increased during the first 35-60 min (e.g., Barber et al. 2007, Keller et al. 2007, Sugita et al. 2005), the Cochran's conclusion may show that duration of the triggered outburst could exceed 35 min. The long ejection is in accordance with the conclusion by Harker et al. (2007) that the best-fit velocity law necessitates a mass production rate that was sustained for duration of 45-60 min after impact.

The size of the region of the DI cloud of essential opacity probably did not exceed 1 km. Our research testifies in favor of a model close to gravity-dominated cratering.

In our estimates of velocities of ejected particles, we analyzed the motion of particles along a distance of a few km. Destruction, sublimation, and acceleration of particles did not affect much our estimates of velocities because we considered the motion of particles during no more than a few minutes. During the considered motion of particles with initial velocities $v_p \geq 20$ m/s, the increase in their velocities due to the acceleration by gas did not exceed a few meters per second (Ipatov & A'Hearn 2011).

The time variations in the rates and velocities of observed particles ejected after the DI impact differed from those found in experiments and in theoretical models. Holsapple & Housen (2007) concluded that these differences were caused by vaporization of ice in the plume and fast moving gas. Their conclusion could be true for the ground-based observations made a few hours after the impact. In our studies of the motion of particles during a few minutes, the greater role in the difference could be played by the outburst triggered by the impact (by the increase of ejection of small bright particles), and it may be possible to consider the ejection as a superposition of the normal ejection and the triggered outburst. The contribution of the outburst to the brightness of the cloud could be considerable, but its contribution to the total ejected mass could be relatively small because the fraction of *small* observed particles among particles of *all* sizes was probably greater for the outburst than for the normal ejecta. Our model of ejection considered only those particles that reached a distance $R \geq 1$ km from the place of ejection. Large regions of saturated pixels in DI images made at time $t$ after impact greater than 110 s prevented us from drawing firm conclusions about the rates of ejection of all particles.

Ipatov and A'Hearn (2011) studied the motion of small particles with velocities greater than the escape velocity (which is equal to 1.7 m/s) at $t$<13 min. These particles constituted a small part of all ejected material. While analyzing DI images, Richardson et al. (2007) and Holsapple and Housen (2007) considered the motion mainly of particles that were ejected with small velocities $v_e$ and fall back on the comet (i.e. they studied the motion of quite different particles than Ipatov and A'Hearn). Richardson et al. (2007) studied the plume base; it was of order 150-350 m in diameter at time 9 to 13 min after the impact. They concluded that >90% of the ejected mass never gets more than a few hundred meters off the surface of the comet, and has been redeposited within 45 min after the impact. Ipatov and A'Hearn (2011) did not analyze the ejection of slow-moving particles and did not make any conclusions based on the particles that were located at $R$<1 km in images made at 1 s<$t$<13 min.

## 4. Peculiarities of the Deep Impact ejection and cavities in Comet Tempel 1

Conditions of ejection of material from Comet Tempel 1 were different from those for experiments and theoretical models. The difficulties in having different gravity, velocities, sizes in laboratory experiments compared to Deep Impact are partly overcome by use of scaling laws involving non-dimensional quantities (see e.g. Housen & Schmidt 1983, Holsapple 1993). The great difference in projectile kinetic energy introduces challenges when scaling the laboratory results to DI conditions, e.g. some materials will vaporize that otherwise would remain in solid or liquid form (Ernst & Schultz 2007). Acceleration of particles by gas is discussed in the previous section.

The fraction of water vaporized at the DI impact is considered to be ~0.2% of the total amount of water ejected (DiSanti et al. 2007). According to Biver et al. (2007), the amount of water released at the DI impact was about 0.2 days of normal activity, but that during the natural outburst on 22-23 June, 2005 was about 1.4 days of normal activity (i.e. was larger than at the DI burst). At the natural outburst, water was in the form of gas, so the outburst was not as bright as the burst after impact. Ipatov and A'Hearn (2011) discussed that a considerable fraction of the brightness of the DI cloud could be due to the triggered outburst (probably except for the first few seconds after impact). The outburst could increase the duration of ejection of material and the velocities of ejected particles and caused the jumps in time variation in the rate of ejection.

A few other differences of the DI ejection from experiments are the following: gravity on the comet (0.04 cm/s$^2$) is much smaller than that on the Earth (9.8 m/s$^2$), and masses of projectiles in experiments were small. Diameters of particles that made the main contribution to the brightness of the DI cloud are considered to be less than 3 μm (e.g. Jorda et al. 2007), and sizes of sand particles in experiments were much larger (~100 μm) than those of the observed DI particles. The observed DI cone of ejected material was formed mainly by small particles, which had higher velocities than larger particles.

For an oblique impact, on the down-range side of the ejecta plume, ejection velocities are higher and particle ejection angles are lowered compared with a vertical impact (Richardson et al. 2007). Besides the outburst, the differences between theoretical estimates (Housen et al. 1983, Holsapple 1993, Holsapple & Housen 2007, Richardson et al. 2007) and observed velocities are partly caused by that in the theoretical model all particles ejected at the same time had the same velocities and were ejected at the same distance from the place of impact. In our opinion, at the same time different DI particles could be ejected with different velocities and at different distances from the center of the crater. The outburst ejection could have come from the entire surface of the crater, while the normal ejection was mainly from its edges.

Analysis of observations of the DI cloud and of outbursts from different comets testifies in favor of the proposition that there can be large cavities, with material under gas pressure, below a considerable fraction of a comet's surface. Internal gas pressure and material in the cavities can produce natural and triggered outbursts and can cause splitting of comets. At a triggered outburst caused by a collision, the duration of the 'fast' outburst (caused by the ejection of material from cavities) can be short because most of the material under pressure can leave the excavated cavity quickly. The most extensive part of the DI triggered outburst took place during about a minute. Duration of some natural outbursts was much longer (weeks or months) because in this case material often moved from cavities through narrow cracks.

Ipatov and A'Hearn (2011) noted that at 1<$t_e$<3 s and 8<$t_e$<60 s the plot of time-variation in the estimated rate $r_{te}$ of ejection of observed material was greater than the exponential line connecting the values of $r_{te}$ at 1 and 300 s. The above features could be caused by the impact being the trigger of an outburst and by the ejection of more icy material. A considerable outburst began at about 8-10 s after DI impact. The beginning of the increase of the outburst could be caused by excavation at $t_e$≈8 s of a large cavity that contained dust and gas under pressure. This cavity could be deep because the duration of the main outburst was about a minute and the direction from the place of ejection to the brightest spot of the cloud of ejected material at 13<$t$<55 s was different from the direction at other moments of time. The beginning of the main excavation of the cavities at $t_{eb}$≈8 s shows that the upper boarders of relatively large cavities were located at about a few meters or more below the surface. With the increase of the crater, more cavities could be excavated. Dust and gas under pressure in pores and small cavities excavated after the impact could also contribute to the outburst ejection, but not to the rays of ejection (excessive ejection in a few directions). N. Gorkavyi noted that the region with cracks through which outburst material could be ejected could grow faster than the crater.

For theoretical models (Holsapple & Housen 2007), radius of a crater is proportional to $t_e^\gamma$, where γ is about 0.25-0.4. Note that $10^\gamma$ is about 1.8-2.5, the diameter $d_f$ of the final DI crater is estimated to be ~100-200 m (e.g., 130-220 m in (Schultz et al. 2007), and not more than 85-140 m in (Richardson et al. 2007)), and the ratio $d_h$ of a crater depth to the diameter is about 1/5-1/3. The duration $T_e$ of the normal ejection is estimated to be not more than 250-550 s (Richardson et al. 2007). These estimates testify in favor of the location of the upper border of the main excavated cavity at a depth $d_{cav}$~5-10 meters. For example, at time $t_{eb}$=8 s, the depth of a crater $d_{cr}=d_f \times d_h/(T_e/t_{eb})^\gamma$=12.5 m for $d_f$=100 m, $d_h$=0.25, $T_e$=80 s, and $(T_e/t_{eb})^\gamma=10^\gamma$=2; for the same data and $T_e$=400 s, $d_{cr}$≈12.5/1.62≈8 m. The distance $d_{cav}$ between the pre-impact surface of the comet's nucleus and the upper border of the cavity could be smaller than $d_{cr}$ because the excavated cavity could be located at some distance from the center of the crater (not below the center). On the other hand, due to cracks caused by the impact, the outburst from the cavity could begin before excavation of the upper border. For small cavities excavated at $t_e$=1 s, the value of $d_{cr}$ (~4-5 m) was smaller by a factor of $8^\gamma$ (i.e. by about a factor of 2) than at $t_e$=8 s. The distances from the upper borders of large cavities to the surface of a comet of about 5-10 m, and sizes of particles inside the cavities of a few microns are in a good agreement with the results obtained by Kossacki and Szutowicz (2011) and discussed in Section 1.

**CONCLUSIONS**

Observations of natural and triggered outbursts from different comets testify in favor of existence of large cavities with material under gas pressure below a considerable fraction of comet's surface. Ejection of observed (mainly with diameter $d$<3 μm) particles from Comet 9P/Tempel 1 after the collision of the Deep Impact module with the comet was greater than theoretical estimates. The difference was caused by the outburst triggered by the impact. The excavation of a relatively large cavity began at $t_e$≈8 s after the impact. This cavity could be deep because the excavation from the cavity could last for at least a few tens of seconds. The beginning of the main excavation of the cavities at $t_e$~8 s shows that the upper boarders of the cavities could be located at about 5-10 meters below the surface. The outburst decreased at ~60 s after the impact. Besides the 'fast' outburst caused by ejection from the cavities, there was a 'slow' outburst ejection, which was similar to the ejection from a 'fresh' surface of a comet and could be noticeable during 30-60 min. The 'fast' outburst with velocities ~100 m/s probably could continue for at least several tens of seconds. It could significantly increase the fraction of particles ejected with velocities ~100 m/s compared with the theoretical models.

**ACKNOWLEDGMENTS**

This research was supported in part by NASA through the American Astronomical Society's Small Research Grant Program.

Reviewed by Dr. Nick Gorkavyi (Greenwich Institute for Science and Technology, VA, USA)